\documentclass[aps,prl,twocolumn,superscriptaddress,nofootinbib,floatfix]{revtex4-1}

\usepackage{amsmath,amssymb}
\usepackage{graphicx}
\usepackage{bm}
\usepackage[colorlinks=true,allcolors=blue]{hyperref}

\newcommand{\Msun}{M_\odot}
\newcommand{\Mbh}{M_{\bullet}}
\newcommand{\Ogw}{\Omega_{\rm GW}}
\newcommand{\tdyn}{t_{\rm dyn}}
\newcommand{\tsink}{t_{\rm sink}}
\newcommand{\tseed}{t_{\rm seed}}
\newcommand{\Myr}{\,{\rm Myr}}
\newcommand{\pc}{\,{\rm pc}}
\newcommand{\Hz}{\,{\rm Hz}}
\newcommand{\mHz}{\,{\rm mHz}}
\newcommand{\Mpc}{\,{\rm Mpc}}
\newcommand{\lnL}{\ln\Lambda}
\newcommand{\fpbh}{f_{\rm PBH}}
\newcommand{\fgas}{f_{\rm gas}}

\begin{document}

\title{The gravitational-wave fingerprint of dynamically assembled \\
primordial black hole cluster seeds in JWST's Little Red Dots}

\author{Juan~Garc\'ia-Bellido}
\affiliation{Instituto de F\'isica Te\'orica UAM-CSIC, Universidad Aut\'onoma de Madrid, Cantoblanco, 28049 Madrid, Spain}

\date{\today}

\begin{abstract}
The James Webb Space Telescope (JWST) has revealed compact, red, overmassive accreting black holes---the so-called ``Little Red Dots'' (LRDs)---in chemically near-pristine hosts at $z\simeq5 - 9$, straining standard heavy-seed models. We show that a population of strongly clustered primordial black holes (PBH) with a broad mass function predicted by a QCD-epoch thermal history naturally realizes the configuration that assembles LRD-scale seeds: an intermediate-mass PBH \emph{nucleus} $\Mbh\sim10^3$--$10^5\,\Msun$ surrounded, within a few parsecs, by a swarm of light ($m\sim30\,\Msun$) PBHs embedded in dense baryonic gas. Gas dynamical friction keeps the loss cone full and lets the core contract, so the swarm sinks and is swallowed on $\tseed\sim10$--$50\Myr$, well inside the cosmic time at $z\sim10$--$15$. Because a heavy nucleus is present \emph{ab initio}, the captures occur at extreme mass ratio $q\sim10^{-4}$--$10^{-2}$: the remnant is retained against gravitational recoil, and each capture reaches the innermost stable orbit under gravitational-wave (GW) emission, radiating $\simeq0.06\,m c^2$ so that the assembly efficiency is $\zeta\simeq0.06$ independent of $\Mbh$. The superposed swarm inspirals form a stochastic background $\Ogw h^2\sim10^{-13}$--$10^{-11}$ with a $\Ogw\propto f^{2/3}$ shape truncated below the gas-decoupling frequency and topped by a ringdown ``comb'' at $f_{\rm ring}(\Mbh)\simeq13\mHz$ for $10^5\,\Msun$ and $\simeq1.3\Hz$ for $10^3\,\Msun$ nuclei at $z_f\simeq12$. The few comparable-mass nucleus--nucleus coalescences are instead individually resolvable LISA/deci-Hz sources. Detection, and discrimination of these signatures from a directly formed PBH seed of the same mass, would identify the LRDs as PBH-nucleus seeded black holes.
\end{abstract}

\maketitle

\textit{Introduction.}---JWST has uncovered an abundant population of faint, broad-line active galactic nuclei at the redshift frontier whose compact morphology and V-shaped continuum define the LRDs~\cite{Harikane:2023zcc,Matthee:2023utn,Greene:2024phl,Kokorev:2024kqm}. They host black holes (BHs) of $M_{\rm BH}\sim10^6$--$10^8\,\Msun$ that are strongly overmassive relative to their stellar hosts~\cite{Maiolino:2023zdu,Inayoshi:2019fun}, though the broad-line masses are actively debated and some may be overestimated~\cite{Greene:2024phl}. The cleanest case is the strongly lensed A2744-QSO1 at $z=7.04$: a dynamical measurement reveals Keplerian rotation about a point mass $M_{\rm BH}\simeq5\times10^7\,\Msun$ with $M_{\rm BH}/M_\star>2$~\cite{Juodzbalis:2025zfs}, in gas that is extremely metal poor, $Z\simeq4.7\times10^{-3}\,Z_\odot$~\cite{Maiolino:2025tih}. We take A2744-QSO1 as the motivating example rather than as representative of the class.

These two facts are jointly awkward for the leading heavy-seed channel. Direct-collapse black holes (DCBHs) form in atomic-cooling halos only where a strong Lyman--Werner flux ($J\gtrsim J_{\rm crit}$) suppresses H$_2$ cooling~\cite{Bromm:2011cw,Begelman:2006db,Inayoshi:2019fun}, which requires a star-forming neighbor and is rare; the early growth is capped by the halo baryon fraction, $M_{\rm BH}/M_{\rm dyn}\lesssim0.1$~\cite{Pacucci:2015rwa}, and the accompanying star formation enriches the gas. A near-pristine, isolated, strongly overmassive BH instead points to a seed assembled gravitationally~\cite{Maiolino:2025tih}.

Primordial black holes~\cite{Hawking:1971ei,Carr:1974nx,GarciaBellido:1996qt} are a natural candidate: they are expected to be strongly clustered at formation~\cite{Clesse:2015wea,Clesse:2016vqa}, and a broad mass function can constitute all of the dark matter (DM)~\cite{Carr:2023tpt}. Crucially, the mechanisms that make PBHs also make them broadly distributed~\cite{Clesse:2017bsw}: the QCD thermal history imprints a peak near the solar mass with a high-mass tail~\cite{Carr:2019kxo}, and quantum diffusion during inflation generates exponentially heavy, non-Gaussian tails and enhanced clustering~\cite{Pattison:2017mbe,Ezquiaga:2019ftu,Ezquiaga:2022qpw}. A dense PBH cluster therefore generically contains a few heavy members: an intermediate-mass PBH (IMBH) \emph{nucleus} ($\Mbh\sim10^3$--$10^5\,\Msun$), bathed in a swarm of light ($m\sim30\,\Msun$) PBHs. Earlier work considered PBHs as SMBH seeds either statistically~\cite{Bean:2002kx,Duchting:2004dk,Carr:2018rid} or by \emph{direct} formation of $\sim10^5\,\Msun$ PBHs with a narrow mass function~\cite{Kawasaki:2012kn}; here the seed is \emph{dynamically assembled}, and we show that this assembly is what makes it loud in GWs.

\begin{figure*}[t]
\centering
\includegraphics[width=0.82\textwidth]{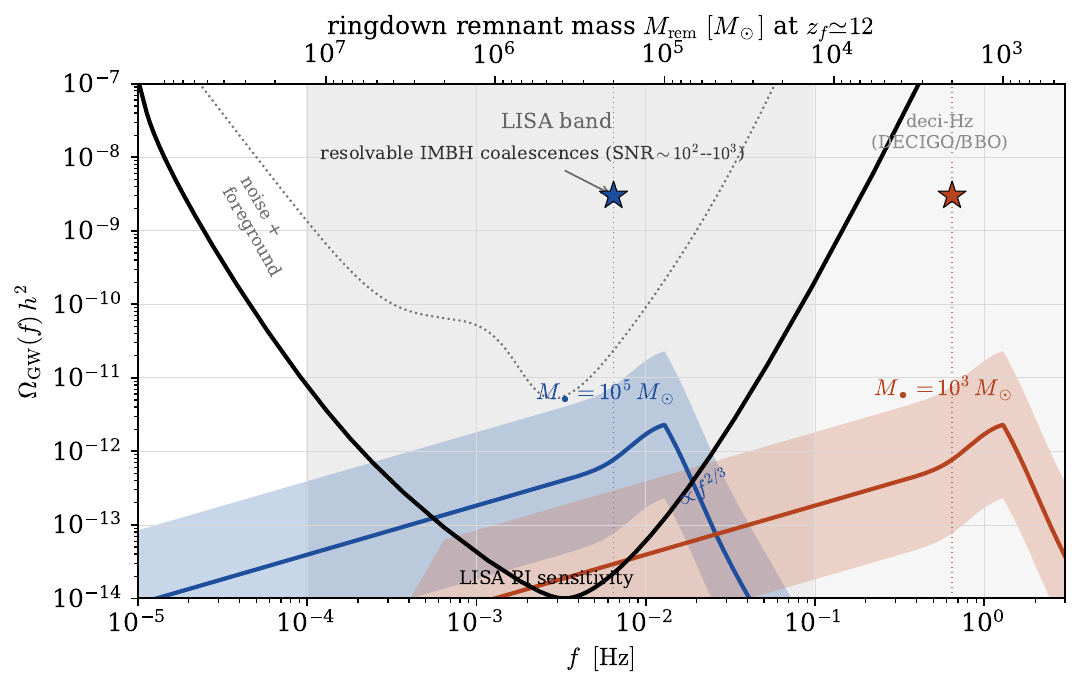}
\caption{\label{fig:lisa} Predicted GW signal from an IMBH nucleus fed by a
light-PBH swarm, for two nucleus masses. Shaded bands: the stochastic
swarm-EMRI background, $\Ogw h^2\propto f^{2/3}$ truncated below the
gas-decoupling frequency and topped by the ringdown comb, spanning
$\rho_{\rm seed}=10$--$10^3\,\Msun\Mpc^{-3}$ [Eq.~\eqref{eq:omeganum}]; solid
curves are the fiducial $\rho_{\rm seed}=100\,\Msun\Mpc^{-3}$. Blue:
$\Mbh=10^5\,\Msun$ (ringdown $\simeq13\mHz$); red: $\Mbh=10^3\,\Msun$
(ringdown $\simeq1.3\Hz$). Stars: the comparable-mass nucleus--nucleus
coalescences, individually resolvable (SNR$\sim10^2$--$10^3$) and removed
before the stochastic search. Black: LISA power-law integrated sensitivity (SNR=1, 4-yr)~\cite{Thrane:2013oya}; dotted: LISA noise $\Ogw h^2$ including the $4$-yr
galactic confusion foreground~\cite{Robson:2018ifk,LISA:2017pwj}. Top axis maps
frequency to remnant mass via Eq.~\eqref{eq:fring}. Detectability of the
background rides the $f^{2/3}$ tail against the foreground; only the optimistic
($\zeta,\rho_{\rm seed}$) corner clears the PI curve, while the resolvable
events are robust.}
\end{figure*}

Embedded in a halo of dense gas, we compute how fast does the swarm is able to build the nucleus into an LRD-scale seed, and what GW signal (background plus resolvable events) does the assembly leave. We find prompt seeding and a distinctive LISA-band signature (see Fig.~\ref{fig:lisa}), and that the nucleus-plus-swarm configuration is free of the recoil, core-collapse-stall and energy-budget difficulties of a runaway of equal-mass PBHs.

\textit{Configuration and seeding time.}---Consider a nucleus $\Mbh$ and $N=\Mbh^{\rm swarm}/m$ light PBHs of mass $m$ within radius $R\lesssim{\rm few}\pc$, embedded in gas of mass $M_g=\fgas\Mbh$ and density $\rho_g=3M_g/4\pi R^3$. A light PBH on an orbit of radius $r$ sinks by the sum of collisionless dynamical friction off the swarm~\cite{Chandrasekhar:1943ws,Binney:2008} and gaseous dynamical friction~\cite{Ostriker:1998fa}; both share the skeleton $t\simeq(\Mbh/m)\,\tdyn/\mathcal{C}$ with $\tdyn=(R^3/G\Mbh)^{1/2}$ and $\mathcal{C}=\lnL$ (collisionless) or $\mathcal{C}=3\mathcal{I}\,\fgas$ (gaseous), $\mathcal{I}\sim\mathcal{O}(1)$ the Mach factor (Supplemental Material~\cite{Note:SM}).

Two features distinguish this from an equal-mass cluster. First, the massive nucleus is present from the start, so that the rate-limiting step of dominant-body formation for a collisionless runaway is absent. For point masses this last process stalls in three-body binary heating and ejection~\cite{Spitzer:1987,Quinlan:1990}, since black holes cannot grow by physical collisions as stars do~\cite{PortegiesZwart:2002iks}. A wide mass spectrum drives Spitzer/gravothermal mass segregation on time scales $t\sim(m/\Mbh)\,t_{\rm rh}\ll t_{\rm rh}$, feeding the nucleus rather than dispersing the core. Second, gas plays a role beyond the drag prefactor: it continuously grinds orbits into the loss cone (avoiding the collisionless loss-cone depletion that throttles capture onto a central BH) and, being dissipative, lets the PBH component \emph{contract}. Because $\tseed\propto R^{3/2}$ is far more sensitive to radius than to any drag coefficient, cooling-driven contraction is the dominant accelerator. We absorb this, and the gravothermal/runaway factors, into a coefficient $\kappa\sim\mathcal{O}(1)$,
\begin{equation}
\tseed\simeq\frac{\kappa}{\lnL}\,\frac{\Mbh}{m}\left(\frac{R^3}{G\Mbh}\right)^{1/2}
=\frac{\kappa}{\lnL}\,\frac{R^{3/2}\Mbh^{1/2}}{m\,G^{1/2}}.
\label{eq:tseed}
\end{equation}
We note that, at \emph{fixed} radius, adding gas mass raises the velocity dispersion and lengthens both drag times as $v^3\propto(1+\fgas)^{3/2}$. Gas helps only through dissipation and contraction, not by deepening the static potential. Numerically, with $\lnL\simeq\ln N$, we find
\begin{align}
\tseed\simeq\,&9\Myr\;\kappa\left(\frac{5.8}{\lnL}\right)\left(\frac{R}{\pc}\right)^{3/2}\nonumber\\
&\times\left(\frac{\Mbh}{10^4\,\Msun}\right)^{1/2}\!\!\left(\frac{m}{30\,\Msun}\right)^{-1}\!\!.
\label{eq:tseednum}
\end{align}
Table~\ref{tab:tseed} lists the two nucleus cases we adopt throughout, $\Mbh=10^3$ and $10^5\,\Msun$. Compact ($R\lesssim1\pc$) configurations assemble $10^3$--$10^5\,\Msun$ seeds in $\sim1$--$50\Myr$, well within the age of the Universe at $z\sim10$--$15$ ($t_H\simeq0.27$--$0.47\,$Gyr); diffuse ($R\gtrsim10\pc$) configurations are excluded by cosmic time, so the scenario self-selects for the compact, strongly clustered cores predicted by enhanced small-scale power~\cite{Clesse:2016vqa,Clesse:2017bsw}. The relaxation clock starts at cluster virialization; since bare compact PBH clusters core-collapse and evaporate on $t_{\rm evap}\lesssim10^8\,$yr~\cite{Spitzer:1987}, the scenario requires cores that reach these densities near the epoch of halo assembly (or are stabilized within dark minihalos) at $z_f \sim 12$.

\begin{table}[t]
\caption{Seeding time, Eq.~\eqref{eq:tseednum}, for $\kappa=1$, and the observed
ringdown frequency of the nucleus, Eq.~\eqref{eq:fring}, at $z_f\simeq12$
($m=30\,\Msun$).}
\label{tab:tseed}
\begin{ruledtabular}
\begin{tabular}{cccccc}
$\Mbh$ & $R$ & $N$ & $\tseed$ & $q=m/\Mbh$ & $f_{\rm ring}$\\
\hline
$10^3\,\Msun$ & $0.3\pc$ & $33$ & $\sim0.5\Myr$ & $3\times10^{-2}$ & $\sim1.3\Hz$\\
$10^3\,\Msun$ & $1\pc$ & $33$ & $\sim3\Myr$ & $3\times10^{-2}$ & $\sim1.3\Hz$\\
$10^5\,\Msun$ & $0.3\pc$ & $3\times10^3$ & $\sim5\Myr$ & $3\times10^{-4}$ & $\sim13\mHz$\\
$10^5\,\Msun$ & $1\pc$ & $3\times10^3$ & $\sim30\Myr$ & $3\times10^{-4}$ & $\sim13\mHz$\\
\end{tabular}
\end{ruledtabular}
\end{table}

\textit{Retention against recoil.}---Because a heavy nucleus feeds on light PBHs, growth proceeds at extreme mass ratio $q=m/\Mbh\lesssim10^{-2}$. The GW recoil of the remnant scales as $v_{\rm kick}\simeq A\,\eta^2$ at small symmetric mass ratio $\eta\simeq q$, with $A\simeq1.2\times10^4\,{\rm km/s}$~\cite{Gonzalez:2006md}. For the $\Mbh=10^5\,\Msun$ nucleus ($q=3\times10^{-4}$) this is a negligible $v_{\rm kick}\sim10^{-3}\,{\rm km/s}$, far below $v_{\rm esc}=(2G\Mbh/R)^{1/2}\simeq29\,{\rm km/s}$ at $R=1\pc$, so retention is automatic. For the $\Mbh=10^3\,\Msun$ nucleus ($q=3\times10^{-2}$) the kick is larger, $\sim10\,{\rm km/s}$, comparable to the bare-core $v_{\rm esc}\simeq3\,{\rm km/s}$ but below the escape speed of the host atomic-cooling halo ($\sim30$--$50\,{\rm km/s}$); such sub-escape kicks are damped and the nucleus re-sinks by gas dynamical friction within a crossing time, so it is retained provided the core sits in a dense halo. This is in sharp contrast to equal-mass assembly, where kicks of $50$--$175\,{\rm km/s}$ (and up to $\sim10^3\,{\rm km/s}$ for spinning superkicks) eject the products and stall the runaway---the classic obstruction to hierarchical IMBH growth~\cite{Campanelli:2007ew,Lousto:2011kp,HolleyBockelmann:2007eh}. The only events that kick significantly are the rare comparable-mass nucleus--nucleus coalescences; these are precisely the individually resolvable sources discussed below, so their ejection would remove a resolvable event, not the background.

\textit{Energy budget.}---A light PBH sinks under dynamical friction until, at the gas--GW handoff radius $a_{\rm h}\sim10^{3}\,G\Mbh/c^2\sim1\,$AU (where the orbital speed exceeds the sound speed and gaseous drag $\propto\rho/v^2$ becomes negligible~\cite{Ostriker:1998fa,Peters:1964zz}), GW emission takes over and drives it through $\sim3$ decades of inspiral to the innermost stable circular orbit (ISCO). Each capture then radiates the ISCO binding energy of the reduced mass,
\begin{equation}
E_{\rm cap}\simeq\varepsilon_{\rm ISCO}\,m c^2,\qquad \varepsilon_{\rm ISCO}\simeq0.057,
\label{eq:Ecap}
\end{equation}
essentially independent of $\Mbh$ for $m\ll\Mbh$. Summed over the $N=\Delta\Mbh/m$ captures that build the nucleus, $E_{\rm GW}=\zeta\,\Delta\Mbh c^2$ with
\begin{equation}
\zeta\simeq\varepsilon_{\rm ISCO}\simeq0.06,
\label{eq:zeta}
\end{equation}
a healthy efficiency set by real GW-driven inspirals. This is the crucial payoff of the nucleus-plus-swarm picture: the same mass ratio that guarantees retention also guarantees that each capture is an inspiral to ISCO rather than a naked plunge. A pure radial plunge would radiate only $\sim0.01\,q\,mc^2$~\cite{Davis:1971gg}, giving $\zeta\sim10^{-5}$; that limit is not realized because gas cannot remove the last few decades of inspiral. The price is that the inspiral tail is \emph{present}, not suppressed. For the rare comparable-mass nucleus mergers a balanced-tree estimate gives $\zeta_{\rm tree}=1-(1-\varepsilon)^{\log_2 N}\simeq0.3$--$0.5$; these dominate the resolvable, not the stochastic, signal. Equation~\eqref{eq:zeta} assumes captures actually reach ISCO rather than being tidally scattered or unbound first, valid for point-mass PBHs.

\textit{Stochastic background.}---Using the cosmic energy-budget formula~\cite{Phinney:2001di,Maggiore:1999vm},
\begin{equation}
\Ogw(f)=\frac{f}{\rho_c c^2}\!\int\! dz\,\frac{R_m(z)}{(1+z)H(z)}\frac{dE_{\rm GW}}{df_s}\Big|_{f_s=(1+z)f},
\label{eq:phinney}
\end{equation}
where $R_m(z)$ is the comoving \emph{capture-rate density}---the number of swarm captures (light-PBH inspirals onto a nucleus) per unit comoving volume per unit source-frame time, in $\Mpc^{-3}\,{\rm s}^{-1}$---and $dE_{\rm GW}/df_s$ is the source-frame GW energy spectrum radiated per capture, $E_{\rm cap}=\int df_s\,(dE_{\rm GW}/df_s)=\zeta\,mc^2$, Eq.~\eqref{eq:Ecap}; the factor $(1+z)^{-1}$ redshifts the source rate to the observer frame. Because assembly is prompt, Eq.~\eqref{eq:tseednum}, $R_m(z)$ is sharply peaked around $z_f$, and its cosmic-time integral fixes the total number of captures per comoving volume, $\int R_m(z)\,dt_s=\int R_m(z)\,[(1+z)H(z)]^{-1}dz=n_{\rm seed}N$, with $N=\Delta\Mbh/m$ the number of captures per nucleus. Carrying out the frequency integral, the background from seeds of comoving mass density $\rho_{\rm seed}=n_{\rm seed}\Mbh$ formed near $z_f$ is
\begin{equation}
\Ogw^{\rm tot}=\frac{\zeta}{1+z_f}\,\frac{\rho_{\rm seed}}{\rho_c}.
\label{eq:omtot}
\end{equation}
The amplitude is set by $\rho_{\rm seed}$ and $z_f$, \emph{not} by $\fpbh$: even for $\fpbh=1$ only $\sim10^{-9}$ of the DM ends in merger-built seeds, so raising $\fpbh$ does not raise this background. What a clustered, all-DM PBH population with a broad mass function provides instead is the \emph{supply} of dense nucleus-plus-swarm cores; we phrase this as a consistency requirement and flag the mapping from the clustering two-point function $\xi(r)$ to $(R,\Mbh,n_{\rm seed})$ as the key remaining calculation. 

For a broad estimate, let us take $n_{\rm seed}\sim10^{-4}$--$10^{-3}\Mpc^{-3}$ and $\Mbh\sim10^3$--$10^5\,\Msun$, $\rho_{\rm seed}\sim10$--$10^3\,\Msun\Mpc^{-3}$ (a few $\times10^{-3}$ of the local So\l tan density~\cite{Soltan:1982vf}, as expected for pre-growth seeds), giving
\begin{align}
\Ogw^{\rm tot}h^2\simeq\,&2\times10^{-12}\left(\frac{\zeta}{0.06}\right)\!\left(\frac{\rho_{\rm seed}}{100\,\Msun\Mpc^{-3}}\right)\!\left(\frac{13}{1+z_f}\right).
\label{eq:omeganum}
\end{align}
Because the assembly is a superposition of GW-driven inspirals, the spectral shape is set by the swarm EMRIs.

\textit{Spectrum: background plus signal.}---A population of circular inspirals gives the canonical $\Ogw\propto f^{2/3}$~\cite{Peters:1964zz}. Here that tail is present but \emph{truncated}: gas hardening removes angular momentum above the GW band, so each inspiral contributes only between the gas-handoff frequency $f_{\rm h}\sim5\times10^{-4}f_{\rm ring}$ and the nucleus ISCO/ringdown. The background per nucleus mass therefore rises as $f^{2/3}$ across $\sim3$ decades, cuts off steeply below $f_{\rm h}$, and is topped by a ringdown ``comb'' at
\begin{equation}
f_{\rm ring}\simeq\frac{1.7\times10^4\Hz}{1+z_f}\left(\frac{\Msun}{M_{\rm rem}}\right)
\label{eq:fring}
\end{equation}
where $M_{\rm rem}$ is the mass of the merger remnant---the nucleus after a capture, $M_{\rm rem}=\Mbh+m-E_{\rm cap}/c^2\simeq\Mbh$ for $q\ll1$, so the comb tracks the growing nucleus mass (we take the $a\simeq0.7$ value; prograde gas-fed accretion can spin the nucleus to $a\gtrsim0.9$ and raise $f_{\rm ring}$ by a further $\simeq1.5$--$2$, an uncertainty in the mass--frequency map). For the two cases, $\Mbh=10^5\,\Msun$ rings at $\simeq13\mHz$ (LISA core) and $\Mbh=10^3\,\Msun$ at $\simeq1.3\Hz$ (deci-Hz), with the $f^{2/3}$ tails sweeping down through the LISA band (Fig.~\ref{fig:lisa}). Because the background is a superposition of discrete, eccentric captures, it is markedly non-Gaussian (``popcorn''), which aids separation from Gaussian foregrounds~\cite{Braglia:2022icu}.

Crucially, the loudest events are not stochastic. The comparable-mass nucleus--nucleus ($10^{3}$--$10^{5}\,\Msun$) coalescences at $z\simeq12$ are gold-plated LISA/deci-Hz sources with SNR $\sim10^2$--$10^3$~\cite{Klein:2015hvg,Ellis:2023owy}, individually resolvable and removable; heavy IMRIs from the high-mass tail of the swarm are likewise resolvable. With $n_{\rm seed}N\sim\mathcal{O}(1)$ ringdowns per $\Mpc^3$ over $\sim10^8\,$yr, the observed burst rate is $\sim10^4$--$10^5\,{\rm yr}^{-1}$ with $\sim10^3\,$s redshifted durations---a duty cycle $\sim0.1$--$1$, confirming that the correct detection statistic is a burst/non-Gaussian search rather than the Gaussian cross-correlation behind the SNR=1 power-law-integrated (PI) curve. We therefore present the resolvable events as a sharper, more falsifiable prediction than the stochastic slope: a few to tens per year of resolvable, eccentric, light-seed coalescences at $z\sim10$--$15$.

\textit{Detectability and the DM fraction.}---The fiducial background peaks at $\Ogw h^2\sim10^{-12}$ near the nucleus ringdown, riding the $f^{2/3}$ tail against the galactic foreground; only the optimistic corner ($\zeta=0.3$, $\rho_{\rm seed}=10^3$) clears the PI curve, so we show the full four-decade uncertainty on $\Ogw^{\rm peak}$ rather than a narrow band. The robust, falsifiable signal is the resolvable population: a directly formed $10^5\,\Msun$ PBH~\cite{Kawasaki:2012kn} carries \emph{no} assembly background and \emph{no} eccentric light-seed coalescences, so the swarm EMRIs and the resolvable nucleus mergers are precisely what discriminates a dynamically assembled seed from a primordial one of the same mass. A deci-Hz observatory (DECIGO/BBO~\cite{Kawamura:2011zz}) captures the $10^3\,\Msun$ ringdown and the high-frequency rollover, providing a multiband lever on the nucleus mass function. At stellar/IMBH masses, for monochromatic spectra, $\fpbh=1$ is excluded by microlensing, the LIGO/Virgo rate and CMB accretion~\cite{Bird:2016dcv,Sasaki:2016jop,Ali-Haimoud:2017rtz} (see however~\cite{Carr:2023tpt}), so the relevant population is a broad mass function whose dense cores supply both the nuclei and swarms; the constituent $m\sim30\,\Msun$ PBH--PBH mergers form the separate, higher-frequency LIGO/ET background~\cite{Sasaki:2016jop,Sasaki:2018dmp,Braglia:2021wwa}.

\textit{Self-consistency.}---The dense gas that enables the mechanism has observable costs: (i)~The orbital energy delivered to the gas during assembly, $E\sim G\Mbh^2/R\sim9\times10^{50}\,$erg for $\Mbh=10^5\,\Msun$, $R=1\pc$, exceeds the gas binding energy for $\fgas\lesssim1$; the very cooling that drives contraction radiates this away, so the gas is regulated, not necessarily unbound, but the deposited energy re-emerges as accretion luminosity. (ii)~The densities required ($n\gtrsim10^6\,{\rm cm}^{-3}$) put the PBHs near Bondi/Eddington accretion, so Bondi growth shortens $\tseed$ only with radiative feedback; the same accretion makes the phase observationally exposed to the unresolved X-ray background and 21-cm limits on accreting PBHs~\cite{Inoue:2017csr}, which we require the fiducial core not to saturate. (iii)~Gas at these densities forms stars unless fragmentation is suppressed; the measured $Z\simeq5\times10^{-3}Z_\odot$ of A2744-QSO1 shows some nearby star formation did occur, and the scenario must specify why the parsec core makes black holes rather than a star cluster---the counterpart of the fine-tuning we ascribed to DCBH---unless these black holes are PBH.

\textit{Conclusions.}---A clustered, broad-mass-function PBH population, as predicted by the QCD thermal history and by quantum diffusion during inflation, naturally forms IMBH nuclei surrounded by light-PBH swarms in very dense halos very early on ($z\gg10$). Gas keeps the loss cone full and contracts the core, so the swarm assembles a $10^3$--$10^5\,\Msun$ seed on a time scale $t\sim1$--$50\Myr$, Eq.~\eqref{eq:tseed}, fast enough to explain the supermassive, chemically pristine LRDs without a Lyman--Werner neighbor or super-Eddington growth. The extreme mass ratio guarantees both retention against recoil and GW-driven captures with $\zeta\simeq0.06$. The assembly can be detected as a non-Gaussian, $f^{2/3}$-truncated stochastic background with a ringdown comb at $f_{\rm ring}(\Mbh)$ $\sim13\mHz$ and $\sim1.3\Hz$ for $10^5$ and $10^3\,\Msun$ respectively, with an amplitude $\Ogw h^2\sim10^{-13}$--$10^{-11}$, plus resolvable, eccentric nucleus and IMRI coalescences that carry the falsifiable weight and discriminate assembled from directly-formed PBH seeds. Fixing $\kappa$, $\zeta$ and the retention budget calls for dedicated $N$-body-plus-hydrodynamics simulations of the gas-embedded, loss-cone-fed nucleus.

\begin{acknowledgments}
The author thanks the BlackTHUNDER team for making the QSO1 data public, and Alexander Kusenko and Sachiko Kuroyanagi for stimulating discussions. He also acknowledges support from the Spanish Research Project PID2024-159420NB-C43 [MICINN-FEDER] and the Centro de Excelencia Severo Ochoa Program CEX2020-001007-S at IFT.
\end{acknowledgments}

\bibliographystyle{apsrev4-1}
\bibliography{PBH_seeds_refs}

\onecolumngrid
\clearpage

\begin{center}
\textbf{\large Supplemental Material:}\\[2pt]
\textbf{\large GW signature of a PBH nucleus fed by a light PBH swarm in dense gas}
\end{center}
\vspace{6pt}

\setcounter{equation}{0}
\setcounter{section}{0}
\setcounter{table}{0}
\renewcommand{\theequation}{S\arabic{equation}}
\renewcommand{\thesection}{S\Roman{section}}
\renewcommand{\thetable}{S\arabic{table}}

We work to order of magnitude; order-unity coefficients ($\kappa,\zeta,\mathcal{I}$) are defined where they appear and are the quantities a dedicated simulation must calibrate. We use $G=c=1$ except in final expressions.

\section{Dynamical-friction sinking, and the role of gas}
\label{sm:df}

\textit{Collisionless channel.}---A body $m$ at radius $r$ in a background of dispersion $v_c=(G\Mbh(<r)/r)^{1/2}$ loses angular momentum to dynamical friction~\cite{Chandrasekhar:1943ws,Binney:2008} at $F_{\rm DF}=4\pi G^2 m^2\rho\,\lnL/v_c^2$. If the friction background is the swarm, $\rho=\rho_\bullet=3\Mbh/4\pi R^3$ and
\begin{equation}
t_{\rm DF}=\frac{v_c^3}{4\pi G^2 m\rho_\bullet\lnL}
=\frac{1}{\lnL}\frac{\Mbh}{m}\,\tdyn,\qquad
\tdyn\equiv\left(\frac{R^3}{G\Mbh}\right)^{1/2}.
\label{eq:tdf}
\end{equation}

\textit{Gaseous channel.}---For transonic motion through gas $\rho_g$, $F_{\rm gas}=4\pi G^2 m^2\rho_g\mathcal{I}/v^2$~\cite{Ostriker:1998fa}, so with $v=(GM/R)^{1/2}$, $\rho_g=3M_g/4\pi R^3$,
\begin{equation}
t_{\rm gas}=\frac{m v}{F_{\rm gas}}=\frac{v^3}{4\pi G^2 m\rho_g\mathcal{I}}
=\frac{1}{3\mathcal{I}}\,\frac{M^2}{m M_g}\,\tdyn .
\label{eq:tgas}
\end{equation}

\textit{Static gas.}---Both times scale as $v^3\propto M^{3/2}$. Setting the dispersion by the \emph{total} mass, $v^2=G\Mbh(1+\fgas)/R$, and adding the rates in parallel,
\begin{equation}
\tsink=\frac{1}{3}\frac{\Mbh}{m}\left(\frac{R^3}{G\Mbh}\right)^{1/2}
\frac{(1+\fgas)^{3/2}}{\lnL+\mathcal{I}\fgas},
\label{eq:tsink}
\end{equation}
so at fixed $R$ and $\Mbh$, $\tsink(\fgas)/\tsink(0)=(1+\fgas)^{3/2}\lnL/(\lnL+\mathcal{I}\fgas)$, which \emph{increases} with $\fgas$ (a factor $\simeq2.4$ at $\fgas=1$ for $\lnL=6$, $\mathcal{I}=1$; even the maximal-drag limit $\mathcal{I}\to\lnL$ gives $\propto(1+\fgas)^{1/2}$). Adding a static gas mass raises the dispersion and weakens both drags faster than the extra channel compensates. Gas is instead useful because it is \emph{dissipative}: it (a)~lets the swarm contract, and since $\tseed\propto R^{3/2}$ this dominates; (b)~refills the loss cone continuously, circumventing the collisionless loss-cone depletion~\cite{Quinlan:1990} that limits capture onto a central mass. Equation~\eqref{eq:tseed} of the Letter absorbs (a) into $\kappa$; the correct late-stage driver is not $t_{\rm DF}(\Mbh)\propto1/\Mbh$ but the growing gravitational-focusing/loss-cone cross-section $\propto\Mbh$, which makes the final captures fast.

\section{Retention against recoil}
\label{sm:kick}

For a non-spinning binary of symmetric mass ratio $\eta=q/(1+q)^2$, the recoil fit of Ref.~\cite{Gonzalez:2006md} gives $v_{\rm kick}=A\,\eta^2\sqrt{1-4\eta}\,(1+B\eta)$ with $A\simeq1.2\times10^4\,{\rm km/s}$, $B\simeq-0.93$. For swarm captures $\eta\simeq q=m/\Mbh$, so $v_{\rm kick}\simeq A q^2$: $\sim10^{-3}\,{\rm km/s}$ at $q=3\times10^{-4}$ ($\Mbh=10^5$) and $\sim10\,{\rm km/s}$ at $q=3\times10^{-2}$ ($\Mbh=10^3$). The first is $\ll v_{\rm esc}=(2G\Mbh/R)^{1/2}\simeq29\,{\rm km/s}$; the second exceeds the bare-core $v_{\rm esc}\simeq3\,{\rm km/s}$ but not the host-halo escape ($\sim30$--$50\,{\rm km/s}$), and is damped by gas friction on a crossing time. Retention against the first-generation nonspinning kick requires $q\lesssim0.05$ (bare core) and is satisfied by both cases once the halo is included. After the first generation the nucleus carries $a\simeq0.7$ and comparable-mass mergers can superkick to $\sim10^3\,{\rm km/s}$~\cite{Campanelli:2007ew,Lousto:2011kp}; these nucleus--nucleus events are the resolvable sources, so their ejection removes a resolvable burst, not the swarm-fed growth.

\section{Radiated energy per capture}
\label{sm:energy}

A test mass inspiralling adiabatically to the Schwarzschild ISCO radiates the ISCO binding energy $\varepsilon_{\rm ISCO}=1-\sqrt{8/9}\simeq0.057$ of $\mu c^2\simeq m c^2$; for $a=0.7$ prograde this rises toward $\sim0.1$. A purely radial plunge from rest radiates instead the Davis--Ruffini--Press--Price energy $E_{\rm plunge}\simeq0.0104\,(m/\Mbh)\,mc^2$~\cite{Davis:1971gg}, i.e.\ $\zeta_{\rm plunge}\simeq0.01\,(m/\Mbh)\ln(\Mbh/m)\sim10^{-5}$. The two differ by whether the final inspiral is GW-driven. Equating the gas-hardening time to the GW inspiral time~\cite{Peters:1964zz} for $m=30\,\Msun$ around $\Mbh=10^5\,\Msun$ puts the decoupling at $a_{\rm h}\sim10^3\,G\Mbh/c^2\sim1\,$AU, leaving $\sim3$ decades of GW-driven inspiral to ISCO. Thus $\zeta\simeq\varepsilon_{\rm ISCO}\simeq0.06$ per unit assembled mass---but the same GW-driven segment produces the $f^{2/3}$ tail, so a suppressed-tail, high-$\zeta$ spectrum cannot both hold. For comparable-mass nucleus mergers, $\zeta_{\rm tree}=1-(1-\varepsilon)^{\log_2 N}\simeq0.35$--$0.5$ for $N=3\times10^2$--$3\times10^4$ ($\varepsilon\simeq0.05$).

\section{Spectral shape}
\label{sm:slope}

Each capture is a GW-driven inspiral with $dE/df_s\propto f_s^{-1/3}$; the superposition over a population gives the standard $\Ogw\propto f^{2/3}$~\cite{Phinney:2001di,Peters:1964zz}, here bounded below by the gas-handoff frequency $f_{\rm h}$ and above by the nucleus ISCO/ringdown, with a ringdown pile-up at $f_{\rm ring}$. From $a_{\rm h}\sim10^3\,G\Mbh/c^2$ and $a_{\rm ISCO}=6\,G\Mbh/c^2$, $f_{\rm h}/f_{\rm ring}\simeq(6/10^3)^{3/2}\simeq5\times10^{-4}$, i.e.\ the tail spans $\simeq3.3$ decades. Eccentricity at plunge shifts power to higher harmonics and broadens the ringdown comb. For a distribution of nucleus masses the combs smear over $1\mHz$--$1\Hz$ and the overlapping tails form a broad plateau.

\section{Ringdown-frequency mapping}
\label{sm:ring}

The $\ell=m=2$ mode of a remnant $M_{\rm rem}$, spin $a$, has $f_{\rm RD}=(c^3/2\pi GM_{\rm rem})[1.5251-1.1568(1-a)^{0.1292}]$~\cite{Berti:2005ys}. For $a=0.7$, $f_{\rm RD}\simeq1.7\times10^4\,(\Msun/M_{\rm rem})\Hz$ (a $62\,\Msun$, $a\simeq0.67$ remnant rings at $\simeq250\Hz$, matching GW150914); redshifting by $(1+z_f)=13$ gives Eq.~\eqref{eq:fring} and the entries of Table~\ref{tab:tseed}. Gas-aligned prograde accretion driving $a\gtrsim0.9$ raises $f_{\rm RD}$ by a further $\simeq1.5$--$2$.

\section{LISA sensitivity and the model of Fig.~1}
\label{sm:fig}

The LISA noise uses the analytic PSD $S_n(f)$ of Ref.~\cite{Robson:2018ifk} plus the $4$-yr galactic confusion foreground, converted via $\Ogw^{\rm noise}h^2=(2\pi^2/3H_0^2)f^3 S_n(f)\,h^2$ with $H_0=2.18\times10^{-18}\,$s$^{-1}$, $h=0.674$; all curves and quoted numbers are $\Ogw h^2$. The black curve is the PI sensitivity~\cite{Thrane:2013oya} for SNR$=1$, $T=4\,$yr. Each nucleus background is modelled as $\Ogw h^2=\Ogw^{\rm pk}h^2\,(f/f_{\rm ring})^{2/3}$ for $f_{\rm h}\le f\le f_{\rm ring}$, a steep cutoff below $f_{\rm h}=5\times10^{-4}f_{\rm ring}$, a ringdown bump at $f_{\rm ring}$, and an $f^{-4}$ rollover above, with $\Ogw^{\rm pk}h^2\in[10^{-13},10^{-11}]$ (fiducial $10^{-12}$). Resolvable nucleus--nucleus coalescences are marked at $f_{\rm ring}(2\Mbh)$.

\begin{table}[h]
\caption{Fiducial parameters and order-unity coefficients.}
\begin{ruledtabular}
\begin{tabular}{lll}
Symbol & Meaning & Fiducial\\
\hline
$m$ & swarm PBH mass & $30\,\Msun$\\
$\Mbh$ & nucleus (seed) mass & $10^3,\,10^5\,\Msun$\\
$R$ & core radius & $0.3$--$1\pc$\\
$z_f$ & formation redshift & $10$--$15$\\
$q$ & capture mass ratio $m/\Mbh$ & $10^{-4}$--$10^{-2}$\\
$\kappa$ & contraction/runaway coeff. & $0.3$--$3$\\
$\zeta$ & GW efficiency ($\simeq\varepsilon_{\rm ISCO}$) & $0.05$--$0.1$\\
$\mathcal{I}$ & gaseous-drag Mach factor & $\mathcal{O}(1)$\\
$n_{\rm seed}$ & comoving nucleus density & $10^{-4}$--$10^{-3}\Mpc^{-3}$\\
$\rho_c$ & critical density & $1.26\times10^{11}\Msun\Mpc^{-3}$\\
\end{tabular}
\end{ruledtabular}
\end{table}

\end{document}